# Onset of rigidity in glasses: from random to self-organized networks

M. Micoulaut[1], P. Boolchand[2] and J.C. Phillips[3]

[1]*Laboratoire de Physique Théorique de la Matière Condensée,*

*Université Pierre et Marie Curie,  Boite 121*

*4, Place Jussieu, 75252 Paris Cedex 05, France*

[2] *Department of ECECS, University of Cincinnati,*

*Cincinnati, OH 45221-0030*

[3] *Department of Physics and Astronomy, Rutgers University*

*Piscataway, NJ  08854-8019*

We review in this paper the signatures of a new elastic phase that is found in glasses with selected compositions. It is shown that in contrast with random networks, where rigidity percolates at a single threshold, networks that are able to self-organize to avoid stress will remain in an almost stress-free state during a compositional interval, an intermediate phase, that is bounded by a flexible phase and a stressed rigid phase. We report the experimental signatures and describe the theoretical efforts that have been accomplished to characterize the intermediate phase. We illustrate one of  the methods used in more detail with the example of Group III chalcogenides and finally suggest further possible experimental signatures of self-organization.

# 1. Introduction

The generic power of Lagrangian methods [1] is familiar from the early days of classical mechanics: the ball rolling down an inclined plane without slipping generates a cycloid, and that same cycloid reappears in apparently unrelated variational problems, such as the tautochrome (a pendulum whose frequency is independent of amplitude) and the brachistochrome (the fastest path between two points for a particle in a constant force field). Similarly, good glasses are known to form in selected regions ("sweet spots") of multicomponent phase diagrams. What is the common or generic element shared by all such networks?

In the early 1980s, it was suggested [2], [3] suggested that a flexible (floppy) network of weakly cross-linked chains will become rigid when the fraction of cross-linking agents reaches a threshold value, and that good glasses could be expected to form near this threshold. In covalent network glasses, one usually expresses the degree of cross-linking in terms of a mean coordination number, $\bar{r}$. In highly cross-linked networks where the connectivity (or $\bar{r}$) is large, there are more constraints than degrees of freedom per atom on average and the structure is stressed rigid (overconstrained or hyperstatic in the language of truss mechanics). At low connectivity, one has a flexible (hypostatic or underconstrained) structure that contains more degrees of freedom than constraints. A floppy to stressed rigid elastic phase transition was predicted to occur at a rigidity percolation threshold value near the network mean coordination number of $\bar{r}$ =2.4. Experimental signatures of this peculiar transition has been found in chalcogenide glasses from Raman scattering [4], Stress relaxation and viscosity measurements [5], vibrational density of states [6], Brillouin scattering [7], Lamb-Mossbauer factors [8], resistivity [9], Kohlrausch fractional exponents [10].

Two new features have emerged in the recent years that have opened new questions and perspectives in this field. First, the underlying nature of the onset of rigidity in glasses has been questioned because two transitions at $\bar{r}_{c1}$ and $\bar{r}_{c2}$ have been found experimentally in a variety of network glasses [11]. These define an intervening region (or intermediate phase, (IP)) between the flexible and the stressed rigid phase, which is surely the "sweet spot" established empirically. In the IP, glasses display some remarkable properties such as absence of ageing [12] or internal stress [13], and selection of isostatically rigid local structures [14]. Moreover, links between IP and protein folding [15], high-temperature superconductors [16] or computational phase transitions [17] have been stressed that go much beyond simple analogies. The understanding of the IP (the generic "cycloid" of strongly disordered phases) is therefore of general interest.

Another striking feature is that these new observations, and more generally concepts and ideas from constraint theory [2,3], are not restricted merely to chalcogenide alloys, as previously applied. Similarly complex elastic phase transitions occur in oxide glasses as well. Specifically, in sodium silicates elastic [18] free energies obtained from Brillouin scattering correlate remarkably well with the computed fraction of floppy modes in the network. In calcium silicates, the onset of conductivity can be related with the breakdown of the stressed-rigid network [19]. Constraint theory has also shed new light on the structure of the most pervasive synthetic material known, window glass (ternary sodium calcium silicates) [20], and even on the most complex network glass family, the borosilicates, which includes pyrex, a quaternary [21].

This suggests that glasses can be classified both from their mechanical and from their thermal properties; the IP was discovered in a mechanical context, and subsequently has been identified from temperature-dependent properties of supercooled melts, and especially the reversibility of the glass transition itself [12]. It suggests that the intermediate phase is a generic feature of disordered networks. It finally encourages the reinvestigation of popular structurally modified glasses within this framework.

This paper attempts to review the theoretical and experimental background of rigidity transitions in the context of these new findings. It highlights the effect of the deviation from network randomness that is called self-organization which originates the intermediate phase. The latter appears when glasses can organize or adapt themselves to lower their free energy at the temperature of glass formation. A simple application for self-organized Group III chalcogenide glasses is presented with one of these theoretical tools. We finally sketch further possible experimental probes as a conclusion.

## 2. Experimental signatures and properties

Early attempts to search for the flexible to rigid elastic phase transition in glassy networks were focused on measurements of bulk elastic constants using ultrasonic echoes [22] and traditional methods [23]. The results were understandably disappointing. It was not until the power of Brillouin scattering [24] was bought to bear on the problem that clear evidence of a gigantic photo-elastic softening of the Longitudinal Acoustic mode in $Ge_xSe_{1-x}$ glasses was observed near the elastic phase boundary. Furthermore, since the underlying length scale of the acoustic excitations exceeds *100 nm* typically, Brillouin scattering in chalcogenide glasses is in essence a *mean-field probe* of the phase transition, and one does not expect non-mean-field effects (such as the IP) to be manifested (see below). In the Brillouin scattering measurements [24] on chalcogenides, the elastic phase transition was manifested as a light-induced effect, and not in the pristine state of glasses. The anticipated increase in elastic constants as a power-law with composition in the rigid phase at low laser excitation was not observed [24]. The conundrum was finally resolved when Brillouin scattering measurements on sodium silicate glasses [18] became available, and showed clear evidence of a power-law variation of elastic constant increments, in harmony with mean-field prediction [3]. In oxide glasses, near-neighbor bond stretching forces far exceed the more-distant neighbor interactions, but the same cannot be said of the lone-pair bearing chalcogenides. In the numerical predictions, the power-law increase in elastic constant increments was deduced for networks stabilized by bond-stretching and bond-bending forces only. Clearly, inclusion of more distant interactions blurs the elastic phase transition.

Raman scattering on chalcogenide glasses has provided remarkably new insights into the onset of rigidity [25]. Compositional variations in the mode frequency squared ($v^2_{CS}$) of the strongly Raman active symmetric stretch of corner-sharing Ge(Se$_{1/2}$)$_4$ tetrahedra (CS mode), in several Ge-based glasses ( Ge-Se [26], As-Ge-Se [27], P-Ge-Se [12]), has revealed two elastic thresholds $\bar{r}_{c1}$ and $\bar{r}_{c2}$ instead of one (Fig. 1). The intervening regions between these two thresholds for the several glass systems in Fig.2, represent IPs. The increasing rigidity of networks possessing a connectivity at $\bar{r} > \bar{r}_{c2}$, was manifested in the anticipated [3] optical elasticity ($v^2_{CS}$) power-laws [26]. Here one must remember that although the scale of the corner-sharing mode frequency, $v_{CS}$, is set by the nearest-neighbour bond-stretching force constants, small variations in the mode frequency, $v_{CS}(\bar{r})$, with increasing connectedness $\bar{r}$ derive from inter-tetrahedral interactions and connectedness of the networks. It is for this reason that Raman scattering probes the elastic behavior of glassy networks at all length scales, and Intermediate Phases are manifested in these experiments. Recently the stress-free nature of IPs was demonstrated in Raman pressure measurements. One has found the threshold pressure, $P_c(x)$ for the CS-mode to shift as a function of external pressure P to vanish [35] in the IP but to monotonically increase both at $\bar{r} > \bar{r}_{c2}$ and at $\bar{r} < \bar{r}_{c1}$. The vanishing of threshold pressures $P_c$s in these Diamond Anvil Cell experiments, is a characteristic feature of crystalline solids, suggesting that glassy networks in IPs exist in a stress-free state or a state of quasi-equilibrium.

But perhaps one of the most striking experimental results to emerge in this area is the recognition that glass transitions become nearly thermally reversing in character in select compositional windows. DSC has been used with great success to establish glass transitions ($T_g$s) over the past four decades. The overshoot of the endothermic excursion in a traditional DSC experiment has more recently been probed with elegance in modulated-DSC experiments [14], a more sensitive variant of traditional DSC. As a glass is heated to $T_g$, changes in structure occur at all length scales as a network softens. Some of these changes are thermally reversing or vibrational in nature, while others are configurational or thermally non-reversing in character. MDSC permits separating these two types of changes by directly deconvoluting the glass transition endotherm into these two components [36]. In these thermal experiments, one finds that the non-reversing enthalpy near $T_g$ largely vanishes in the same range of chemical compositions that were identified above as IPs in Raman scattering (Fig. 1). Fig.2 depicts in a schematic fashion the compositional windows across which glass transitions become thermally reversing in character- *reversibility windows*. The existence of reversibility windows was not predicted by theory. The observation adds to the physical picture of IPs in showing that there are profound thermal consequences on the nature of glass transition when networks self-organize in IPs [14] as discussed in the next section.

Taken together, Brillouin scattering, Raman scattering, modulated-DSC provide compelling evidence for the opening of stress-free (isostatically rigid) intermediate phases between flexible (floppy) and stressed- rigid ones in disordered networks.

## 3. From random to self-organized networks

The nature of the intermediate phase and the two boundaries as described above at $\bar{r}_{c1}$ and $\bar{r}_{c2}$ have been characterized from theories that follow approaches developed for initial constraint theory [2,3]. All come to the conclusion that networks which self-organize to avoid stress from additional cross-linking agents will display a rigidity transition at low $\bar{r}$ and a stress transition at high $\bar{r}$. In the mean-field approach or in random networks where self-organization does not take place, both transitions coalesce into a single one. This furthermore happens to be the case in glasses where photo-induced excitations produce the collapse of the intermediate phase [24].

In random networks when only nearest-neighbour forces dominate between N atoms, small displacements from the equilibrium structure can be described by a harmonic Kirkwood – Keating potential [37] that contains bond-stretching and bond-bending terms without any further consideration of more weaker forces such as dihedral or Van der Waals forces. Thorpe [3] analyzed the solution of the eigenmodes of such kind of networks constrained only by these local forces and showed that the corresponding dynamical matrix (of dimensionality 3N) has 3N degrees of freedom. If low-frequency (floppy) modes which diagonalize the matrix are present, then the system can be deformed with a low cost in energy and is flexible.

The total number of zero-frequency modes can be computed from Maxwell constraint counting as initially derived in Ref. [2]. For the reader's convenience, we recall here the construction. For a r-coordinated atom, the enumeration of bond-stretching and bond-bending forces (presumed intact for all atoms) gives respectively *r/2* and *(2r-3)* constraints. The latter is obtained by considering a two-fold atom which has only one single angular constraint. Each additional bond onto this atom needs the definition of two additional constraints. (If some bond-bending constraints are broken [as for O in $SiO_2$], the corresponding count is modified appropriately.) The total number of constraints per atom is therefore:

$$n_c = \frac{\sum_{r \geq 2} n_r \left[ r/2 + (2r-3) \right]}{\sum_{r \geq 2} n_r} \tag{1}$$

where $n_r$ is the concentration of atoms with coordination r and $N = \sum_{r \geq 2} n_r$ represents the total number of atoms of the network. The fraction *f* of floppy modes is then given by: *f=3-$n_c$* that can be simply rewritten as a function of the network mean coordination number $\bar{r} = \frac{1}{N} \sum_{r \geq 2} r n_r$ :

$$f = 2 - \frac{5}{6} \bar{r} \tag{2}$$

and vanishes when $\bar{r}$ reaches the magic value of $\bar{r}_c = 2.4$ which can be attained by various compositional combinations.

The vanishing of *f* defines a *single* transition between a flexible and a rigid phase, when constraint counting is performed at a global level without any consideration of the presence of correlated fluctuations that may permit to delay the onset of rigidity when $\bar{r}$ is steadily increased. Similarly, the eigenmode analysis such as described by Thorpe [3] is realized on networks with bonds being rewired at random. Equations (1) and (2) only rely on the coordination number of the local structures which are defined by the macroscopic concentration. Both approaches therefore define a mean-field result where neither typical length-scales nor any spatial correlations of the emerging elastic phases (stress free, stressed) are involved.

The discovery of the Intermediate Phase and the definition of two phase boundaries shows that the previous descriptions (dynamical matrix, constraint counting) can not account for the special observations depicted in Figs. 1 and 2. Theory has therefore brought new refinements and ingredients in the initial approach. These theoretical methods follow the lines defined for the random case however, i.e. they use either the eigenmode analysis or the floppy mode fraction from constraint counting but along new schemes, self-organized networks, that are reviewed next.

**A The "pebble game algorithm"**

Using a graph-theoretical approach, Thorpe and co-workers have been developing an algorithm that takes into account the non-local characteristics of rigidity percolation and allows to calculate the number of floppy modes, to locate over-constrained regions and identify rigid clusters for simple bar-joint networks [38]. This improved algorithm, the Pebble Game, *uniquely* decomposes the network into rigid clusters and determines all stressed rigid regions and permits to consider rigidity in any type of network, from random bond (atoms of a network are randomly connected) to networks with closed ring structures. Details of the method can be found in Refs. [39]-[40].

In the case of simulated self-organized networks, the addition of bonds in a low-connected flexible network will be accepted by the algorithm only if this leads to isostatically rigid clusters. This means that any additional bond leading to the creation of stress in a cluster will be rejected in the procedure. The latter can be realized as long as the bond density is not too high. Indeed, increasing the cross-link density leads to a stiffening of the structure and will finally induce the percolation of rigidity (*a rigidity transition*) that is unstressed (at $\bar{r} = 2.375$ in Fig. 3). Additional bonds now contribute to the occurrence of stressed rigid clusters that finally percolate at a second transition located at the *stress transition* in Fig. 3 at $\bar{r}$ =*2.392*. Apart from the improvement of the computational algorithm, the major difference with the initial analysis of rigidity percolation [3] is that the selective bond addition leads to a new phase, an intermediate phase, which is mainly isostatically rigid, i.e. stress free.

### B Constraint counting and cluster approximations

An alternative framework [41] has been proposed. It uses size increasing cluster approximations (SICA) combined with constraint counting that enables to describe the Intermediate Phase and to go beyond mean-field-rigidity. As already mentioned above, in mean-field rigidity constraint counting is performed either on local structures (e.g. *GeSe$_{4/2}$* tetrahedra and *Se* chain fragments) or on the macroscopic concentration ($n_c$=*2+5x* in *Ge$_x$Se$_{1-x}$*) which leads in all cases to a single rigidity transition satisfying *f=0* (e.g. *x=0.20* in *Ge$_x$Se$_{1-x}$*). One has therefore to find an alternative way to go beyond the local structure approach (or in other words to generate some medium range order) and to apply constraint counting algorithms. Cluster approximations appear to be helpful in trying to build in a self-consistent way structures that contain medium range order such as cyclic (ring) structures and to generate self-organized networks. The method was first introduced (in a slightly different manner) to describe amorphous semi-conductors [42] and has been first applied to *Ge$_x$Se$_{1-x}$* in the context of the onset of rigidity and the nature of the intermediate phase [41].

The SICA construction starts from the usual mean-field treatment of rigidity, where the probabilities of the local structures are derived from the macroscopic modifier concentration (e.g. *x* in *Ge$_x$Se$_{1-x}$*). This corresponds to a basic level (*l=1*) out of which size increasing clusters (*l=2*, *l=3*, ...) are generated and their probabilities $p_i^{(l)}$ (*i=1...N$_l$*) computed. The latter depend on the basic probabilities (l=1), statistical and Boltzmann factors (see example in Table I or in [41],[43]). Here $N_l$ represents the total number of created clusters at a given step *l*. Next, one applies constraint counting algorithms on these clusters by enumerating for each bond-stretching and bond-bending forces, following the counting introduced by Thorpe [3]. This leads to the number of floppy modes of the network given by:

$$f^{(l)} = 3 - n_c^{(l)} = 3 - \frac{\sum_i n_{c(i)} p_i^{(l)}}{\sum_i N_i p_i^{(l)}} \tag{3}$$

where $n_{c(i)}$ and $N_i$ are respectively the number of mechanical constraints and the number of atoms of the cluster with probability $p_i^{(l)}$. An example of such counting is provided in Table I . It shows that the bonding types lead to clusters with a well-defined mechanical character: flexible, isostatically rigid (stress-free) and stressed rigid.

Self-organization is achieved as follows. Starting from all possible flexible clusters in which stressed rigid dendritic connections are absent and increasing the connectivity (or $\bar{r}$ ), one constructs all possible clusters by avoiding a dendritic (non cyclic) stressed rigid connection. This amounts to have selection rules in the cluster construction with growing step *l*. Then, one can investigate at which composition the network will have a vanishing of the number of floppy modes $f^{(l)}$, following the usual definition of the rigidity transition. However, because of the self-organization and the absence of the stressed rigid connections, stress can not propagate with increasing $\bar{r}$. Once the rigidity transition is bypassed, these selection rules can hold for a while but only up to a certain point in connectivity where the network will not be able to avoid stressed rigid dendritic clusters any more. This defines the stress transition in rigidity theory and can not be found from a mean-field treatment. An simple example of such a construction is provided below.

**C Other self-organized networks**

Additional investigation of self-organized networks have been reported by Barré et al. [44] who considered a random bond network able to adapt its free energy. Mousseau and co-workers [45] have also shown recently that self-organization with equilibration on diluted triangular lattices would lead to an intermediate phase. Finally, we mention that Molecular Dynamics studies have failed up to now in considering compositionally driven self-organized glasses such as network glasses ($Ge_xSe_{1-x}$, $As_xSe_{1-x}$, etc…): progress has been made in finding accurate force-fields ([46]), but the problem of finding good space-filling starting configurations unrelated to amorphous Si [3] remains open. Signatures of intermediate phases have been only found numerically from pressure-induced rigidity [47] in simulated $GeO_2$ and $SiO_2$. The question of whether Molecular Dynamics can describe these new observed effects for network glasses is unanswered at this stage.

### D A case study: self-organized Group III chalcogenides

We illustrate in this section how self-organized rigidity affects the behaviour of Group III chalcogenides such as $Ga_xSe_{1-x}$ or $B_xS_{1-x}$. Surprisingly, these compounds have been rarely studied in sulphide and selenide systems with respect to their Group III composition (see however [48]) although some of them form rather easily glasses at the stoichiometric composition ($B_2S_3$, $Ga_2S_3$,…). Studies on Group III tellurides have been reported in the context of optical switching applications [49] but because of the coordination change of tellurium in these systems, we restrict the present study to selenides and sulphides for which the coordination number of the chalcogen is always 2. Also, the coordination number of the Group III Element is well defined, in contrast with Group V chalcogenides where several coordination numbers can exist over selected ranges of composition [12], [14]. We illustrate the construction with the example of $B_xS_{1-x}$. At the first approximation step ($l=2$), the model can be solved analytically as described next.

The considered basic SICA at l=1 units are a $S_{1/2}$ specie and a $BS$ cluster with respective probability ($1$-$p$) and $p$. The choice of these basic units is made to allow that some connections between them (at step $l>1$) can yield isostatically rigid clusters which are optimally constrained ($n_c=3$). This happens to be the case when the two aforementioned units are connected together (Table I) whereas two connected $S_{1/2}$ species are flexible and a $B_2S_2$ cluster stressed rigid (overconstrained). For the latter, several possibilities can occur, depending on how the two basic ($l=1$) clusters connect together (Fig. 5) as one create either a homopolar $B$-$B$ bond or a cyclic ring structure (edge-sharing triangles) that are observed [50] in the stoichiometric $B_2S_3$. This has incidentally some implications on the constraint counting as such structures share some extra constraints due to ring closure that have to be removed [3]. Furthermore, the ring made of two $BS_{3/2}$ triangles connected by edges is a 2D planar ring and is therefore stressed rigid.

Application of equ.(3) in the present case gives for the fraction of floppy modes per atom f at the basic SICA level $l=1$:

$$f^{(1)} = 3 - \frac{2 + 11p}{1 + 3p} = 1 - \frac{5}{2}x \qquad (4)$$

as $x=2p/(1+3p)$. This leads to the location of the rigidity transition estimated from the vanishing the fraction of floppy modes given by $x=2/5=0.4$ (i.e. $B_2S_3$), a result that can be obtained also from a direct counting on the formula $B_xS_{1-x}$.

At the next step ($l=2$) one can now produce some self-organization to avoid stress, i.e. allowing only a certain type of agglomeration between the basic units. With increasing boron content $x$, this amounts to select in the cluster construction only structures that do not contain a connection between two $BS$ basic units (Table I) unless a ring is formed. This means that one allows stress to nucleate in small rings that have a low connectivity (Fig. 4). With this simple construction, one will accumulate isostatically ($n_c=3$) $BS$-$S_{1/2}$ rigid clusters only with increasing $x$, starting from a flexible chalcogen-rich cluster. Mathematically, the whole construction is closed by writing a normalization condition (the sum of all probabilities at a given step must be equal to one) and a specie conservation law, which can be written in the flexible and stressed rigid phase respectively as:

$$p = \frac{1}{2}\left[p_{BS_{3/2}} + 2p_{B_2S_2-ring}\right] = \frac{x}{2 - 3x} \qquad (5)$$

and

$$p = \frac{1}{2}\left[p_{BS_{3/2}} + 2p_{B_2S_2-ring} + 2p_{B_2S_2}\right] = \frac{x}{2 - 3x} \qquad (6)$$

Solving equation (5) and (6) permits to evaluate the involved Boltzmann factors as a function of the Group III concentration and leads to:

$$e_{flex} / e_{iso} = \frac{15x - 6 + 2xe_r}{4(2x-1)} \qquad (7)$$

$$e_{stress} / e_{iso} = \frac{15x - 6 - 2xe_r}{5x} \qquad (8)$$

The computation of the number of floppy modes (application of equ (3)) in the flexible phase leads in this case to:

$$f^{(2)} = 3 - \frac{2p_S + 7.5 p_{BS_{3/2}} + 13 p_{B_2S_2-ring}}{p_S + 2.5 p_{BS_{3/2}} + 4 p_{B_2S_2-ring})} \qquad (9)$$

with the probabilities given by Table I and the Boltzmann factor $e_{flex}/e_{iso}$ given by equ. (7). At some place, one will satisfy $f^{(1)}=0$ which defines the rigidity transition composition that is given by the combination of equs (7) and (9) and reduces to the rigidity threshold $x_{rigidity}=0.40$.

The present selection rules of the SICA construction can be fulfilled only up to a certain point in composition $x_{stress}$ beyond which these stressed rigid non-cyclic *BS-BS* connections cannot be avoided any more. This happens when the energy gain associated with a flexible connection becomes infinite, i.e. when $e_{flex}/e_{iso}$ vanishes, and it defines a second transition given by :

$$x_{stress} = \frac{6}{15 - 2e_r} \qquad (10)$$

The location of the stress transition depends on the factor $e_r$, i.e. on the rate of small stressed rigid rings existing in the structure. So one can see from this simple example that the width of the intermediate phase will depend on how two stressed rigid *BS* clusters will be able to connect themselves together. If ring tendency is favourable (large Boltzmann factor $e_r$), than the width will be large as the system is able to avoid the higher coordinated stressed rigid clusters more longer.

The probability of finding flexible, isostatically rigid clusters and stressed rigid clusters is shown in Fig. 5. As one can see, the growth of edge-sharing connected triangles (i.e. the increase of $e_r$ in this model) leads to an increase of the width of the Intermediate Phase with a stress transition that is shifted to higher connectivities. It shows furthermore that the first-order jump usually observed experimentally [26]-[27] at the stress transition becomes weaker as the fraction of small rings is increased as noticed from the weakening of the kink observed at $x_{stress}$ in the probability of stressed rigid clusters in Fig. 5. This is a feature that has been also obtained in Group IV chalcogenides [41] and on simulated networks using the Pebble Game Algorithms [40]. Expanding the construction to higher steps $l$ is under consideration.

## 4. Space-filling properties and Self-Organization

Another interesting property of the intermediate phase seems to be the tendency of a network to densify the structure. The configurational enthalpy and entropy of glasses is associated with an exponentially small fraction of free-particle configuration space, and as such is inaccessible to numerical simulations using even the largest and fastest computers (which operate in polynomial time). The general rule for glass structures are that they are dominated locally by the local structure of the high-temperature phase (the stable phase near the glass transition temperature), but have slightly lower density (by a few %). Thus in silica the dominant local structure is that of cristobalite, not quartz (the low-T phase). The *Si* atoms in cristobalite occupy a *Si* lattice, and tricks are known for generating *a-Si* networks [50]; these are readily converted to glassy silica by inserting O atoms between nearest neighbour *Si* atoms [51]. In this way good starting points are known for simulations of silica, but this method is not general, and equally good results have so far not been obtained for other network glasses.

The tricks that generate the *a-Si* structure involve ring deformations [52], and it has long been known that rings are the first small cluster with a significantly non-polynomial deviation from "random" packing of smaller clusters (such as tetrahedra). True, quartz and cristobalite have the same small-ring size (six), and quartz achieves its higher density using a much larger unit cell than cristobalite. Thus it appears that by focussing on two factors, the network stress as minimized through matching constraints with degrees of freedom, and average ring size, one should be able to obtain two conditions and determine the composition of window glass [20], the most common synthetic material in current use. The success of this approach depends on having molecular clusters ($SiO_2$, $Na_2O$, and $CaO$) with similar molar volumes. With increasing differences in molar volumes, the composition range of the reversibility window shifts (Fig. 2).

Usually, the compactness optimum is invoked (see e.g. [53, 54]) from the simple argument that a mechanical stability [2] is reached with the vanishing of equ. (2), i.e. at $\bar{r} = 2.4$. This would link the volume contraction with tight bonding and shorter bond lengths when the number of constraints matches exactly the number of degrees of freedom. Inspection of Fig. 6 and Figs 1 and 2 shows that the location of the molar volume minimum and the vanishing of the non-reversing heat $\Delta H_{nr}$ are related in several chalcogenide glasses (e.g. Ge-Se). It suggests that the space-filling tendency is a consequence of the self-organization of a stress-free network. The probability of finding stress-free (isostatically rigid) clusters is indeed maximum at the point where the molar volume is minimum (Fig. 6).

Space-filling tendency on these topological arguments holds as long as the size of the atoms is about the same. In this respect, the issue of space-filling tendency in alkali and alkaline earth silicates [58] where self-organization has now been described [18] and how it relates to nearest neighbour interaction models and simulations remains unanswered at this stage.

## 6. Conclusion

We have reviewed the experimental signatures of a new intermediate phase that appears between the flexible and stressed rigid phase in network glasses with increasing cross-link density. Theoretical methods suggest that this phase appears only when a network is able to self-organize in order to avoid the stress arising from the cross-linking elements. This leads to the definition of two compositional thresholds defining a rigidity transition when the network becomes rigid but stress free, and a stress transition at which the network can not avoid the percolation of stressed rigid clusters more longer. Random networks therefore do not display intermediate phases and both transitions coalesce.

Additional experimental signatures of this peculiar phase should be observable from conductivity measurements. Following the mechanical nature of the network backbone, diffusion of conducting species such as e.g. alkali ions should display different regimes. The Arrhenius activation energy for conduction depends on contributions that usually contain a Coulombic part (the energy required to create a carrier) and a strain part (the energy that is needed to create a "doorway" of conduction in the network). Both contributions should display different behaviours in flexible, intermediate and stressed electrolytes. Recently, the onset of conduction has been found in calcium silicates in the flexible phase [59], in relationship with thresholds seen from optical methods [43]. Obviously, fast ionic conductors in which the conductivity is much larger than in this silicate system, should display even more pronounced signatures of the three phases.

Finally, the question of how diffraction may or may not detect self-organization remains an unanswered question at this stage although compositional changes in the Ge-Se system have been studied with X-ray and neutron scattering [60]. A further open question deals with the relationship between self-organization and medium-range order elements that can be probed from diffraction. As theoretical methods have been developing topological approaches to describe intermediate phases and since connectivity plays here the key role, the neutron Bhatia-Thornton structure factor of number-number correlations (that focuses more on topology than on chemistry) may be the appropriate function [61] to study these new elastic effects in detail and detect any possible signature.

**Acknowledgements**

It is a pleasure to acknowledge ongoing discussions with A.R. Bishop, B. Goodman, M. Malki, N. Mousseau, G.G. Naumis, P. Simon, K. Trachenko, M.F. Thorpe. LPTMC is Unité Mixte de Recherche du Centre National de la Recherche Scientifique (CNRS) n. 7600. The present research has been supported by a joint collaboration grant between CNRS and NSF.

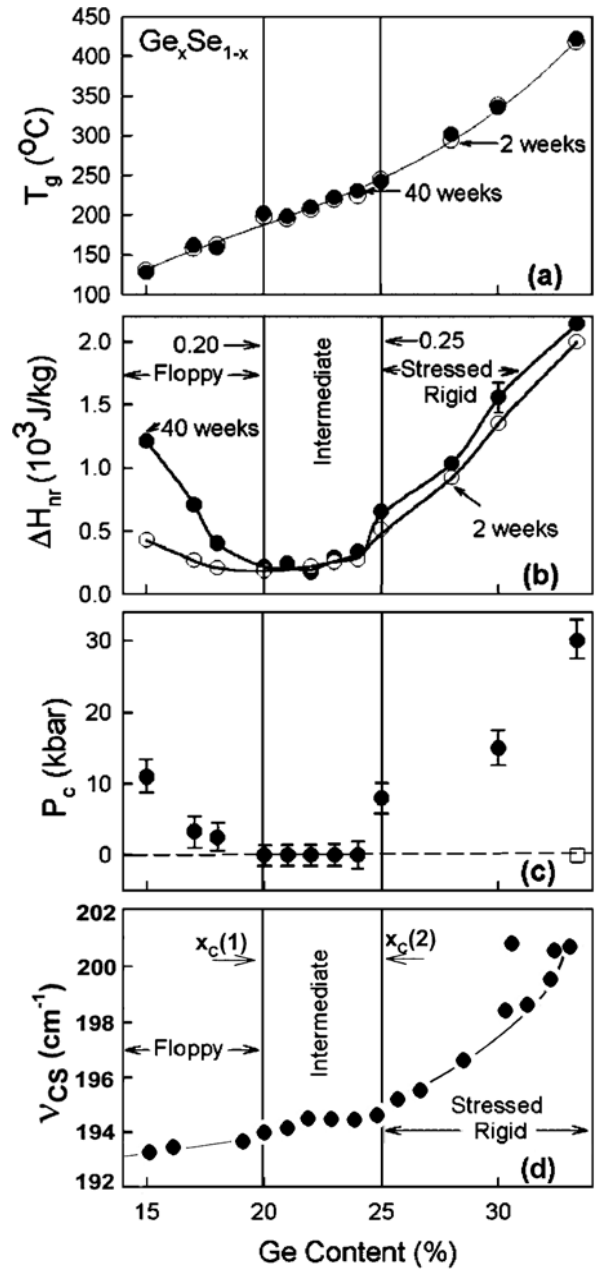

Fig. 1: Experimental signature of the Intermediate Phase in $Ge_xSe_{1-x}$ between $x=0.20$ and $x=0.25$ seen from differential scanning calorimetry (MDSC) in the Non-reversing enthalpy $\Delta H_{nr}$ (panel b), pressure window (panel c) and symmetric stretch mode frequency $\nu_{CS}$ of corner-sahring tetrahadra (panel c) from Raman spectroscopy. Note also weak ageing effects in the window (panel b). After Refs. [26], [35].

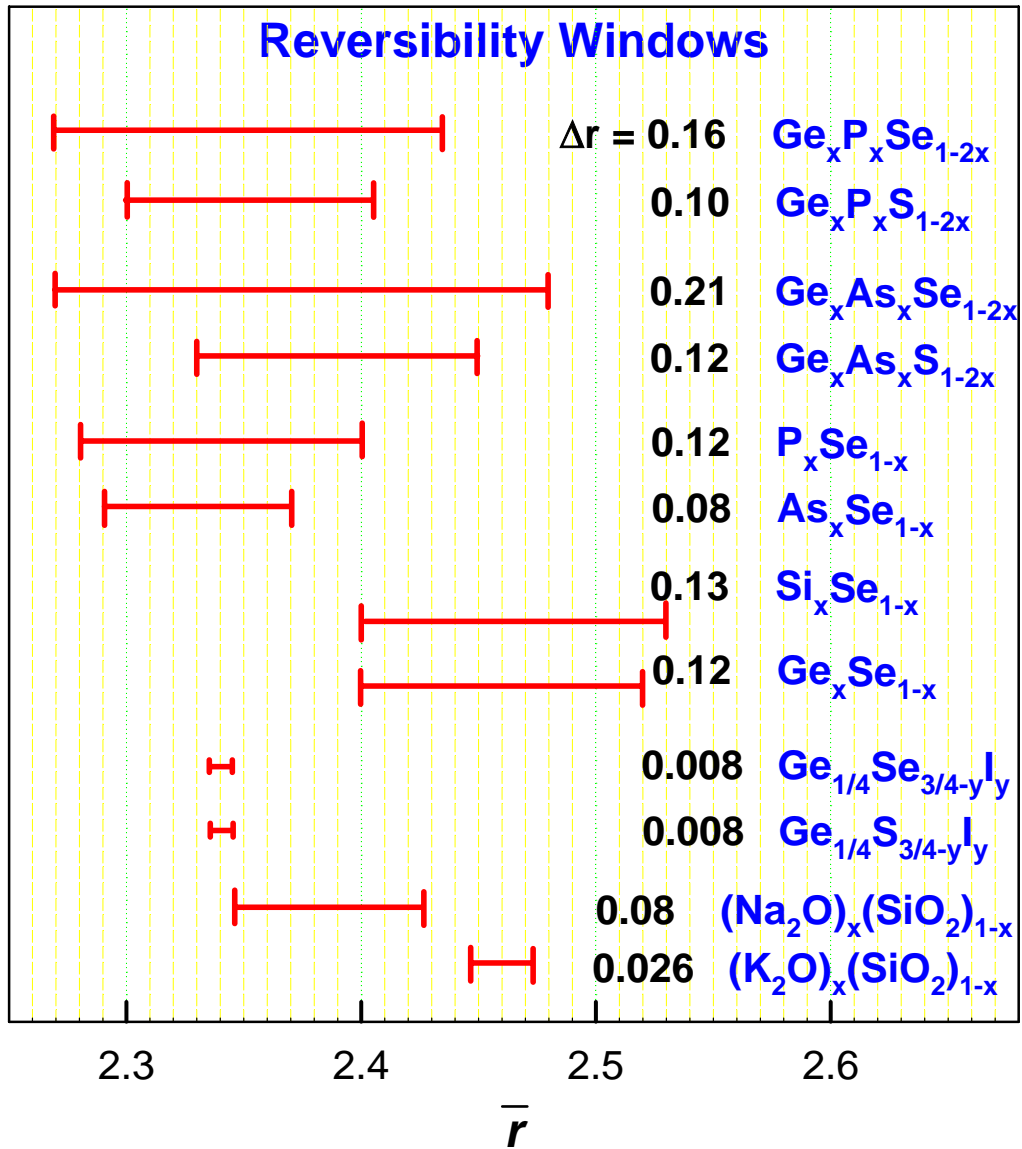

Fig. 2 Width of the intermediate phase for a selected number of glasses. Data on *Ge_xP_xSe_{1-2x}* [12], *Ge_xP_xS_{1-2x}* [28], *GexAs_xSe_{1-2x}* [29], *Ge_xAs_xS_{1-2x}* [30], *P_xSe_{1-x}* [31], *As_xSe_{1-x}* [14], *Si_xSe_{1-x}* [11], *Ge_xSe_{1-x}* [26], *Ge_{1/4}Se_{3/4-y}I_y* [32], *Ge_{1/4}S_{3/4-y}I_y* [33], *xNa_2O-(1-x)SiO_2* [18] and *xK_2O-(1-x)SiO_2* [34].

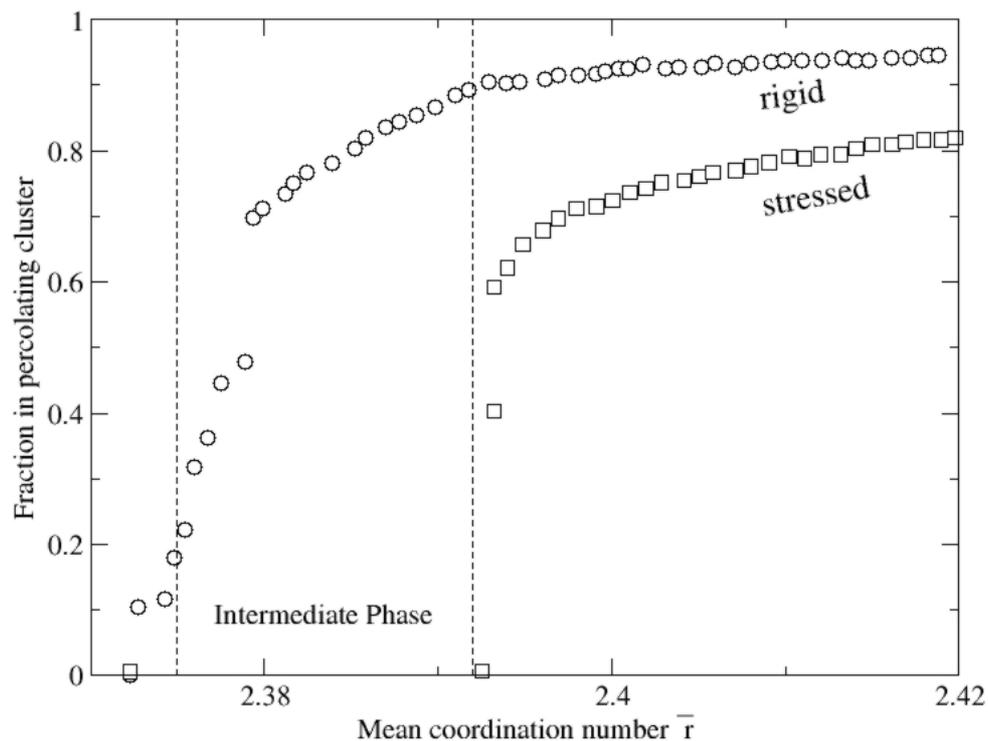

Fig.3 : Solution of the « *Pebble Game* » on a diluted triangular lattice. Fraction of sites on isostatically rigid and stressed rigid percolating cluster (open circles and boxes) in a self-organized network as a function of the network mean coordination number. The intermediate phase which is rigid but unstressed, exists for *2.375< $\bar{r}$ <2.392*. For random networks, both transitions coalesce. After Ref. [39]-[40].

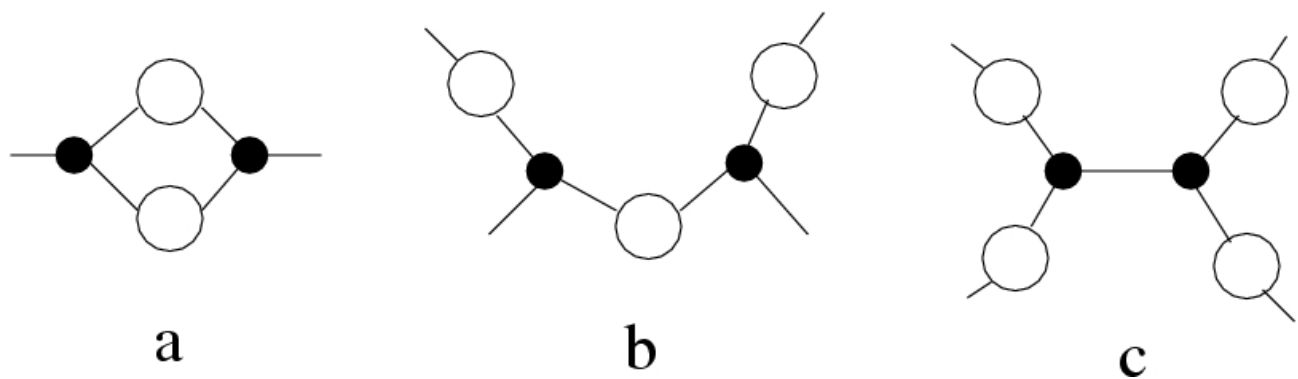

Fig. 4 : Three possible configurations for a stressed rigid $B_2S_2$ cluster at step l=2. The connectivity of these three clusters is respectively 2, 4 and 4. Stress can therefore propagate more easily with cluster b and c.

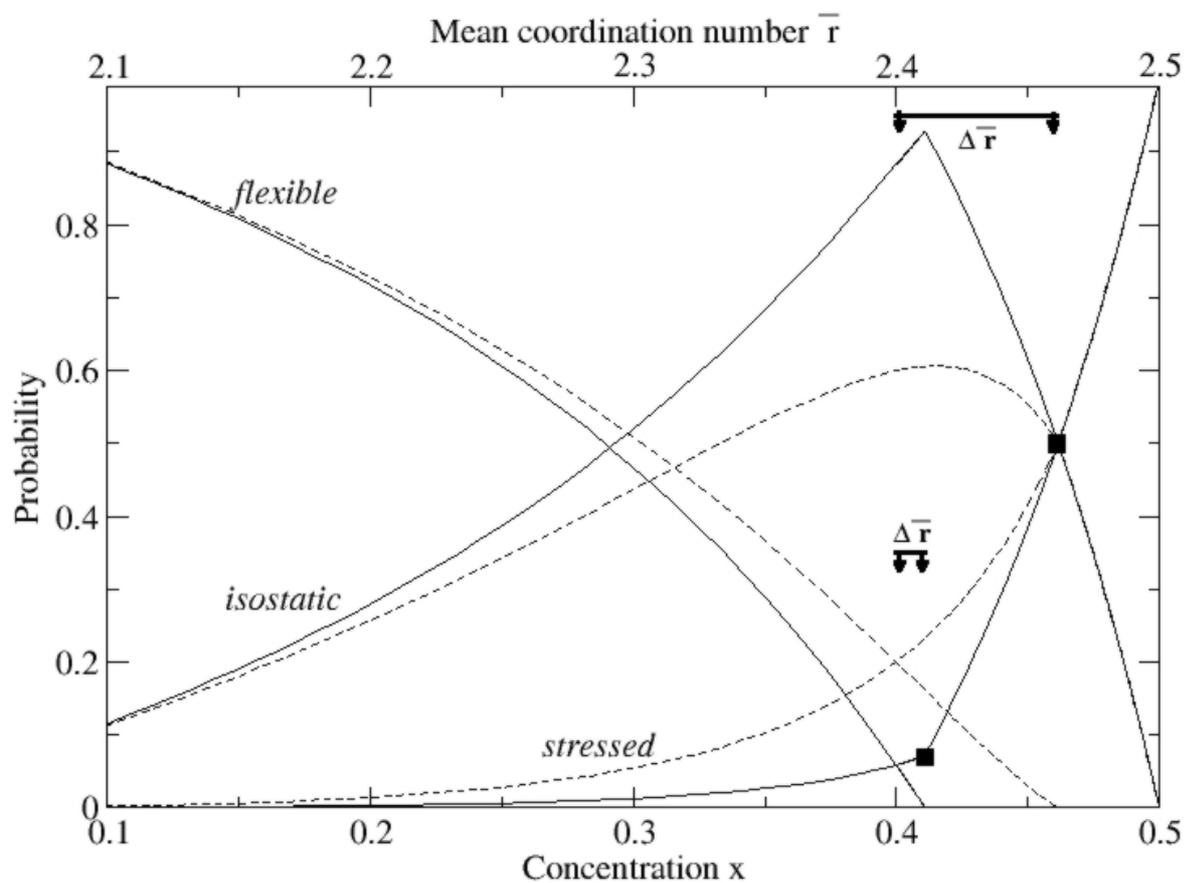

Fig. 5: Probability of finding flexible, isostatically rigid and stressed rigid clusters in Group III chalcogenides (e.g. $B_xS_{1-x}$) as a function of the concentration $_x$ (lower x-axis) or mean coordination number $\bar{r}$ (upper x-axis). Solid line: $e_r=0.2$. Broken line: $e_r=1.0$. The filled boxes indicate the stress transition. The horizontal bars with arrows serve to define the two intermediate phases of respective width $\Delta\bar{r}=0.011$ and $\Delta\bar{r}=0.061$.

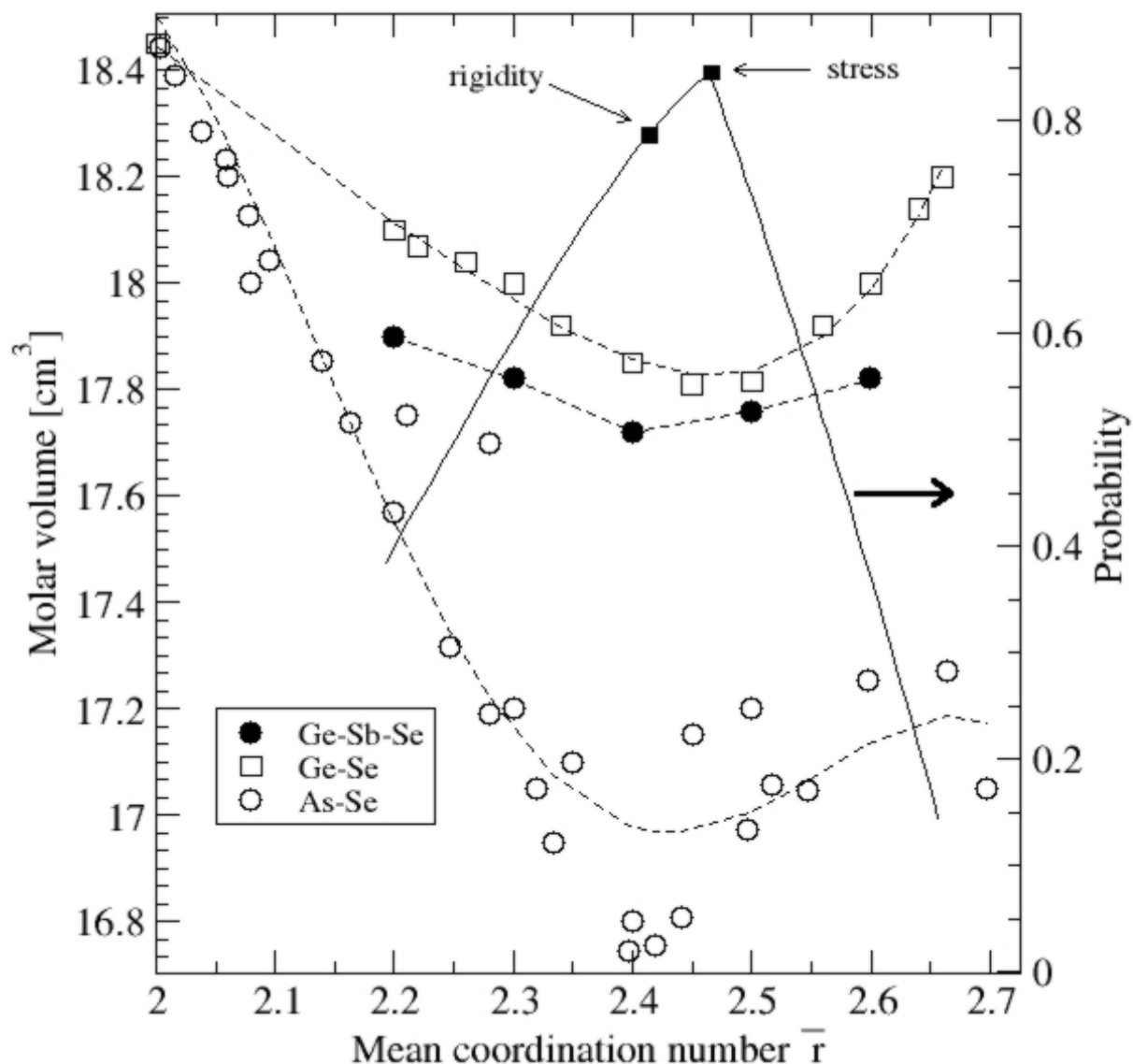

Fig. 6: Molar volumes in some selected chalcogenides: Ge-Sb-Se [55], Ge-Se [56] and As-Se [56], [57]. The broken lines are guides for the eye. Right axis: probability of finding stressed rigid SICA clusters in Group IV chalcogenides with edge-sharing units (from Ref. [41]). The filled boxes indicate the location of the rigidity and stress transition in the model.

| Size l | Cluster | Probability | Number of constraints per atom nc | Mechanical nature |
|--------|---------|-------------|-----------------------------------|-------------------|
| 1 | $S_{1/2}$ | $1-p$ | 2 | Flexible |
| | $BS$ | $p$ | 3.25 | Stressed rigid |
| 2 | $S$ | $2(1-p)^2 e_{flex}$ | 2 | Flexible |
| | $BS_{3/2}$ | $4p(1-p)e_{iso}$ | 3 | Isostatically rigid |
| | $B_2S_2\text{-ring}$ | $2p^2 e_{ring}$ | 2.75 | Stressed rigid |
| | $B_2S_2$ | $5p^2 e_{stress}$ | 3.25 | Stressed rigid |

Table I: Cluster generated at the first SICA steps with their number of constraints, their unrenormalized probabilities and their mechanical nature. Note that the probabilities involve four Boltzmann factors, $e_{flex}$, $e_{iso}$, $e_{ring}$ and $e_{stress}$ following the mechanical (flexible, isostatic, stressed) or topological (dendritic, ring) nature of the connections.